\documentclass[aps,prd,amsmath,amssymb,preprintnumbers,onecolumn,11pt,nofootinbib]{revtex4}
\usepackage[a4paper,left=20mm,right=20mm,top=25mm,bottom=25mm]{geometry}
\usepackage{amsmath,amssymb}
\usepackage{mathrsfs}
\usepackage{graphicx}
\usepackage{color}
\usepackage{subfigure}
\usepackage{fancyhdr}
\usepackage{multirow}
\usepackage{tikz}
\usepackage{feynmp}
\usepackage{mathpazo}
\usepackage{bbold}

\usepackage{epsfig}
\usepackage{amsfonts}
\usepackage{bm}
\usepackage{xcolor}

\definecolor{CP3}{cmyk}{0,0.88,0.77,0.40}

\begin{document}

\title{\Large \color{CP3} Preheating in an inflationary model with disformal coupling}
\author{Khamphee Karwan$^{1}$}
\email{khampheek@nu.ac.th} 
\author{Phongpichit Channuie$^{2}$}
\email{channuie@gmail.com}
\affiliation{\footnotesize $^{1}${The Institute for Fundamental Study \lq\lq The Tah Poe Academia Institute\rq\rq, \\Naresuan University, Phitsanulok 65000, Thailand}}
\affiliation{\footnotesize $^{2}${School of Science, Walailak University, Nakhon Si Thammarat, 80160 Thailand}}

\begin{abstract}
In this work, we investigate the preheating mechanism in a disformally-coupled inflationary model where the scalar field $\phi$ (which is the inflaton field) naturally coupled to another matter field $\chi$ induced by the disformal transformation. In the present scenario, novel derivative interactions mixing the kinetic terms of the two fields emerge inherently. We start by deriving the evolution of the background system when the back reaction on the background field is neglected. We examine the particle production due to parametric resonances in the models and find in Minkowski space that the stage of parametric resonances can be described by the Mathieu equation. Interestingly, we discover that broad resonances in our models can be typically achieved. Finally, we compare our results with previously studied model with derivative couplings.
 \\[4mm]
\end{abstract}

\maketitle

\section{Introduction}
The big bang cosmology has been greatly successful in explaining the evolution of the universe. However, despite its success, it falls short in describing underlying nature of some fundamental physics problems. For example, three of the leading unsolved mysteries include cosmic inflation, dark matter and dark energy, whose underlying descriptions remain yet unknown. In attempting to solve some (all) of the problems, cosmologists and gravitational physicists have sought out the possible modifications of the Einstein gravity on the cosmic scales.  Among many scenarios, theories that include an additional scalar field provide compelling candidates for alternative theories of gravity. In general, the modifications may introduce new gravitational degrees of freedom in addition to the metric tensor and can be in principle described by a scalar-tensor theory of gravity \cite{Fujii2009}.

However, in any non-degenerate theory that time derivative of fundamental dynamical variable is higher than second order, there exist a linear instability, or Ostrogradski's instability. Such instabilities whose Hamiltonian is not bounded from below lead to negative norm states or negative energy states which in higher derivative theories are terminologically called \lq\lq ghost-like\rq\rq. Besides, theories with ghostlike degrees of freedom provide inconsistencies with the experimental tests. To avoid the presence of the Ostrogradski instability, the Euler-Lagrange equations have to be at most second-order. Until recently, it turns out that very successful models of modified gravity can be described in terms of a class of Horndeski's scalar-tensor theory \cite{Horndeski:1974wa}. In spite of the existence of the derivative interactions, the Horndeski's theory is known as the most general scalar-tensor theory in four dimensions with one scalar degree of freedom whose equations of motion are kept up to second order in time and spatial derivatives and therefore they are deprived of the Ostrogradski's instability.

The Horndeski's theory has been investigated so far for various cosmological purposes, e.g. dark energy \cite{Chow:2009fm,Silva:2009km,Deffayet:2010qz,DeFelice:2011hq}, screening mechanisms \cite{Nicolis:2008in,Burrage:2010rs,DeFelice:2011th,Kimura:2011dc,Koyama:2013paa} and also inflation \cite{DeFelice:2011zh,Kobayashi:2010cm,Burrage:2010cu}. In attempting to generalizations, we may consider a scenario that the scalar field is directly coupled to the matter sector. In this case, matter does not follow geodesics associated with the gravitational metric $g_{\mu\nu}$ but instead with another metric $\bar{g}_{\mu\nu}$ that in the simplest situation they are related via ${\bar g}_{\mu\nu} = \Omega^{2}(\phi)g_{\mu\nu}$ which is known as the conformal transformation (or coupling) \cite{Clifton:2011jh}. Here it is worth noting that the matter frame metric $\bar{g}_{\mu\nu}$ can be constructed by the purely gravitational one $g_{\mu\nu}$ and only one scalar field $\phi$, but not by the derivatives of the scalar field itself. The gravity and matter frames, $g_{\mu\nu}$ and $\bar{g}_{\mu\nu}$, are often referred to as the Einstein and Jordan frames, respectively.

As the most general vistas of the scalar-tensor theory, the matter frame metric which is conformally constructed from the gravitational metric and the scalar field itself can also be further generalized by adding the derivatives of the scalar field. Since the higher derivative terms in the equations of motion give rise to ghost-like instabilities associated with the Ostrogradski's theorem, the simplest viable case is to keep only the first order derivatives of the scalar field, {\it viz}.;
\begin{eqnarray} \label{disfor}
{\tilde g}_{\mu\nu} = C(X,\,\phi)g_{\mu\nu} + D(X,\,\phi)\phi_{,\mu}\phi_{,\nu}\,\,,
\end{eqnarray}
where $\phi_{,\mu} =\nabla_{\mu}\phi$ is the covariant derivative of the scalar field associated with the gravity frame $g_{\mu\nu}$ and $X:= -\frac{1}{2}\phi_{,\mu}\phi^{,\mu}$. Note that the above relation is often called the disformal transformation. In the present work we will restrict ourself to the simpler argument in which the disformally-transformed gravitational sector belongs to a subset of the Horndeski class \cite{Horndeski:1974wa,Bettoni:2013diz}, namely;
\begin{eqnarray} \label{disfort}
{\tilde g}_{\mu\nu} = C(\,\phi)g_{\mu\nu} + D(\,\phi)\phi_{,\mu}\phi_{,\nu}\,\,.
\end{eqnarray}
For $D=0$, Eq(\ref{disfort}) reduces to the ordinary conformal transformation and its consequences have been well-established, at least when it depends on $\phi$ only, while when $C=1$, $D$ features the pure disformal transformation. Regarding the scalar-tensor theory with the disformal coupling, there are plenty of compelling applications in the cosmological problems. For instance, the authors of Ref.\cite{Koivisto:2008ak,Zumalacarregui:2010wj,vandeBruck:2015ida} showed that the disformal transformations are very useful to devise models of the dark sector. In particular, inflationary models have also been investigated in contexts of the disformal coupling \cite{Kaloper:2003yf,vandeBruck:2015tna}.

However, another crucial issue for successful models of inflation is the (pre)reheating mechanism. In this work, we anticipate to investigate this mechanism for an inflationary model with the presence of a disformal coupling between two scalar fields.
The reheating and preheating mechanisms for inflationary models in which the couplings between inflaton and matter fields are induced from conformal transformation have been studied in \cite{conformal1, GarciaBellido:2008ab}.
The reheating process in inflationary scenario with disformal coupling between inflaton and other scalar field has been discussed in \cite{Kaloper:2003yf}.
The preheating process due to the direct kinetic coupling between inflaton and other scalar field has been investigated in \cite{ArmendarizPicon:2007iv,Myung:2016eut}.

The paper is organized as follows: In Sec.(\ref{s2}), we discuss the evolution of the background system and then we quantify inflationary trajectories by specifying a particular choice of free functions, $C$ and $D$. In Sec.(\ref{sec3}), we will investigate the particle production due to parametric resonances in a model of inflation in which the disformally-transformed gravitational sector belongs to a subset of the Horndeski class Eq.(\ref{disfort}). For this model of inflation, the scalar field $\phi$ (which is the inflaton field) naturally coupled to another scalar field $\chi$ induced by the disformal transformation. We compare our results with previously studied model with derivative couplings and summarize our findings in the last section. 

\section{Scalar-tensor theory with the disformal coupling}
\label{s2}
We will here start by deriving the set of transformations relating background fields of a disformally coupled scenario. The full action for our model is described by the following action (in Planck units):
\begin{eqnarray} \label{genaction}
{\cal S} = \frac{1}{2}\int d^{4}x\sqrt{-g}R - \int d^{4}x\sqrt{-g}\Big[\frac{1}{2}g^{\mu\nu}\phi_{,\mu}\phi_{,\nu} + {\cal U}(\phi)\Big] - \int d^{4}x\sqrt{-{\tilde g}}\Big[\frac{1}{2}{\tilde g}^{\mu\nu}\chi_{,\mu}\chi_{,\nu} + {\cal V}(\chi)\Big]\,.
\end{eqnarray}
Bearing in mind that the model can be seen as a scalar-tensor theory in the Einstein frame coupled to matter fields that propagate on a physical metric denoted by ${\tilde g}_{\mu\nu}$. The relation between the two-frame metrices is given by the disformal transformation, defined in Eq.(\ref{disfort}). The transformation features how the physical metric for the matter in the action for this theory is disformally related to the gravitational metric. The stress energy momentum tensors in the Jordan frame is defined by
\begin{eqnarray} \label{JF}
{\tilde T}^{\mu\nu} := \frac{2}{\sqrt{-\tilde{g}}}\frac{\delta(\sqrt{-\tilde{g}}{\tilde {\cal L}}_{\rm m})}{\delta {\tilde g}_{\mu\nu}}\quad{\rm with}\quad{\tilde {\cal L}}_{\rm m} \equiv \frac{1}{2}{\tilde g}^{\mu\nu}\chi_{,\mu}\chi_{,\nu} + {\cal V}(\chi)\,,
\end{eqnarray}
for which we can commonly impose a perfect fluid description by defining a Jordan frame energy density, ${\tilde \rho}$, four-velocity ${\tilde u}^{\mu}$, and pressure, ${\tilde P}$:
\begin{eqnarray} \label{rhop}
{\tilde T}^{\mu\nu} = \left({\tilde \rho} + {\tilde P}\right){\tilde u}^{\mu}{\tilde u}^{\nu} +{\tilde P}{\tilde g}_{\mu\nu}\,.
\end{eqnarray}
In the Einstein frame, we also have 
\begin{eqnarray} \label{EF}
T^{\mu\nu} := \frac{2}{\sqrt{-g}}\frac{\delta(\sqrt{-\tilde{g}}{\tilde {\cal L}}_{\rm m})}{\delta g_{\mu\nu}}\,.
\end{eqnarray}
Using the relations in Eq.(\ref{JF}) and Eq.(\ref{EF}), a map between the two objects can readily be derived to obtain:
\begin{eqnarray} \label{EFJF}
T^{\mu\nu} = \sqrt{\frac{\tilde g}{g}}\frac{\delta {\tilde g}_{\alpha\beta}}{\delta g_{\mu\nu}}{\tilde T}^{\alpha\beta} = C^{3}\gamma {\tilde T}^{\mu\nu}\,.
\end{eqnarray}
where the disformal scalar $\gamma$ given below in Eq.(\ref{gammag}) parameterizes the relative contribution of the disformal factor. Using a disformal relation to the relatively simple form of Eq.(\ref{disfort}), we can rewrite the action in terms of $g_{\mu\nu}$ to yield \cite{vandeBruck:2015tna}
\begin{eqnarray} \label{pureg}
{\cal S} = \frac{1}{2}\int d^{4}x\sqrt{-g}\Big[R - g^{\mu\nu}\phi_{,\mu}\phi_{,\nu} - 2{\cal U} - \frac{C}{\gamma}\left(g^{\mu\nu}\chi_{,\mu}\chi_{,\nu}+ 2C{\cal V}\right) + \gamma D \left(\phi_{,\sigma}\chi^{,\sigma}\right)^{2}\Big]\,,
\end{eqnarray}
where the arguments of $C,\,D,\,{\cal U}$ and ${\cal V}$ are understood, and for convenience we have defined the parameter $\gamma$ as
\begin{eqnarray} \label{gammag}
\gamma^{2} = \left(1 + \frac{D}{C}g^{\mu\nu}\phi_{,\mu}\phi_{,\nu}\right)^{-1}\,.
\end{eqnarray}
It is worth noting that the authors of Ref.\cite{DeCross:2015uza} show that the resonance behaviors in the non-minimally-coupled paradigms during the preheating state are sensitive to the oscillation of the background field(s). Early in the oscillation phase, the conformal stretching of the Einstein-frame potential makes the background field behave like a minimally coupled field in a quadratic potential, $V(\phi)=m^{2}\phi^{2}/2$, instead of a quartic potential, $V(\phi)=\lambda\phi^{4}/4$. 

However, in our present investigation, the relation between the two-frame metrices is governed by the disformal transformation, not just a conformal one. By comparision, our Einstein-frame action (\ref{pureg}) is accidentally corroborated with the one in Ref.\cite{DeCross:2015uza} after employing the conformal transformation, saying $D=0$. Here in this specially case, $\gamma=1$ and $C^{-1}\equiv f(\phi^{I}),\,i=1,2$. Along the inflaton-field direction, it is noticed by comparing Eq.(3) in Ref.\cite{DeCross:2015uza} and Eq.(\ref{pureg}) in our model that the Einsten-frame potential takes the form
\begin{eqnarray} 
{\cal U}(\phi) \equiv V(\sigma(\phi^{1})) = \frac{1}{4f^{2}(\sigma(\phi^{1}))}{\tilde V}(\sigma(\phi^{1}))\,,
\end{eqnarray}
where ${\tilde V}(\phi^{1})$ is the Jordan-frame potential, $\phi$ and $\sigma$ are canonically normalized fields but $\phi^{1}$ is not. Since the potential in the Einstein frame in Ref.\cite{DeCross:2015uza} (see also in Refs.\cite{DeCross:2016fdz,DeCross:2016cbs,Kaiser:2015usz}) strongly depends on the conformal factor $f(\sigma(\phi^{1}))$ containing the non-minimal coupling between the two fields, then the oscillation behavior of the inflaton field in their case is highly sensitive to the non-minimal coupling. In contrast, in case of our present scenario, the Einstein-frame potential is independent of the non-minimal coupling. As a result, the oscillation behavior of the inflaton field in our case does not depend on the non-minimal coupling featuring the minimally coupled nature. Moreover, the potential in the Einstein-frame action including the one of our present model can be in principle derived from the nontrivial forms of any potential in the Jordan-frame action.
Nevertheless, if we start from the Jordan frame by supposing that the Lagrangian of the scalar field $\phi$ takes a canonical form,
the disformal transformation will generate the terms like $(\phi_{,\mu}\phi^{,\mu})^2$ in the Einstein frame action.
These terms appear although the field $\chi$ has no contribution to the dynamics of the universe.
Hence, these terms could largely alter the oscillation behaviour of the inflaton field during preheating compared with our case,
and could modify the feature of inflation compared with that in  \cite{vandeBruck:2015tna}.
However, in this work, we  concentrate on the preheating process of model based upon the disformal inflationary scenario investigated in \cite{vandeBruck:2015tna},
so that we will not consider this situation.

In our case by neglecting the back reaction on the background field, we will see that the dynamics of the inflaton field is independent of $C$ and $D$. In order to obtain the field equations, we perform the variation of the action (\ref{pureg}) with respect to the metric $g_{\mu\nu}$ to yield
\begin{eqnarray} \label{fieldeq}
{\cal G}_{\mu\nu} = {\cal T}_{\mu\nu} = T^{(\phi)}_{\mu\nu} + T^{(\chi)}_{\mu\nu}\,,
\end{eqnarray}
where $T^{(\phi)}_{\mu\nu} $ is the usual energy-momentum tensor for a minimally-coupled scalar field and $T^{(\chi)}_{\mu\nu}$ involves the cross terms due to the disformal coupling:
\begin{eqnarray} \label{Tphimunu}
T^{(\phi)}_{\mu\nu} = -\left(\frac{1}{2}g^{\alpha\beta}\phi_{,\alpha}\phi_{,\beta} + {\cal U}\right)g_{\mu\nu} + \phi_{,\mu}\phi_{,\nu}\,,
\end{eqnarray}
and 
\begin{eqnarray} \label{Tmunu}
T^{(\chi)}_{\mu\nu} &=& -\left[\frac{C}{\gamma}\left(\frac{1}{2}g^{\alpha\beta}\chi_{,\alpha}\chi_{,\beta} + C{\cal V}\right) - \frac{1}{2}\gamma D\left(\phi_{,\sigma}\chi^{,\sigma}\right)^{2}\right]g_{\mu\nu} + \frac{C}{\gamma}\chi_{,\mu}\chi_{,\nu} - 2\gamma D\left(\phi_{,\sigma}\chi^{,\sigma}\right)\chi_{,(\mu}\phi_{,\nu)}\nonumber\\&&+ \left[\gamma C \left(\frac{1}{2}g^{\alpha\beta}\chi_{,\alpha}\chi_{,\beta} + C{\cal V}\right) + \frac{\gamma^{3}D^{2}}{2C}\left(\phi_{,\sigma}\chi^{,\sigma}\right)^{2}\right]\phi_{,\mu}\phi_{,\nu}\,,
\end{eqnarray}
and $T^{(\chi)}$ is the trace of $T^{(\chi)}_{\mu\nu}$. Here we obtain the equation of motion for a field $\phi$ as
\begin{eqnarray} \label{ephig}
\Box\phi -{\cal U}' -Q = 0\,,
\end{eqnarray}
and $Q$ is given by a rather complicated expression:
\begin{eqnarray} \label{qg}
Q = \gamma^{2}\left(T^{(\chi)\mu\nu}\nabla_{\mu}\Big[\frac{D}{C}\phi_{,\nu}\Big] - \frac{1}{2C}\Big[C'T^{(\chi)} + D'T^{(\chi)\mu\nu}\phi_{,\mu}\phi_{,\nu}\Big]\right)\,,
\end{eqnarray}
where primes denote derivatives with respect to the field $\phi$. Likewise, varying the action with respect to the field $\chi$ we also obtain its equation of motion:
\begin{eqnarray} \label{echig}
\Box\chi -C{\cal V}' &-& \frac{\gamma^{2}}{2}\left[(\gamma^{2} - 3)\frac{C'}{C} - (\gamma^{2} - 1)\frac{D'}{D} \right]\left(\phi_{,\sigma}\chi^{,\sigma}\right)\nonumber\\&-& \gamma^{2}\frac{D}{C}\left[\phi^{,\mu}\phi^{,\nu}\nabla_{\mu}\chi_{,\nu} + \left(\phi_{,\sigma}\chi^{,\sigma}\right)\Box\phi\right] + \gamma^{4}\frac{D^{2}}{C^{2}} \left(\phi_{,\sigma}\chi^{,\sigma}\right)\phi^{,\mu}\phi^{,\nu}\nabla_{\mu}\phi_{,\nu} = 0\,.
\end{eqnarray}
According to Ref.\cite{vandeBruck:2015tna}, we will consider models where the couplings are given by
\begin{eqnarray} \label{DC}
C(\phi) &=& C_{0} e^{\alpha \phi}, \cr
D(\phi) &=& D_{0} e^{\beta \phi}.
\end{eqnarray}
Here, we have used four parameters, $C_{0},\,\alpha,\,D_{0}$ and $\beta$ to describe our coupling functions.
However, an interesting special case of this parametrisation is when $\alpha=\beta=0$ and the couplings become constants, $C(\phi) = C_{0}$ and $D(\phi) = D_{0}$. Inspired by string theory, the scale $D_{0}$ can be identified with the (inverse) tension of the 3-brane ($T_{3}$) whose an associated disformal mass scale $m_{D}$ is given by $m_{D}=D^{-1/4}_{0}=T^{1/4}_{3}$ \cite{vandeBruck:2015tna}.

For model of inflation, the authors of Ref.\cite{vandeBruck:2015tna} discovered that fields with sub-Planckian initial conditions can drive large amounts of inflation in the presence of a large enough disformal coupling. Here they used $\alpha=\beta=0$, $C_{0} = 1$ and $D_{0}=5.0 \times 10^{21}$ which corresponds to a mass scale of $m_{D} = 3.76 \times 10^{-6}$.
Since $\gamma$ takes an extremely large value, and remains fairly constant (still much greater than unity) until the end of inflation, it is claimed in \cite{vandeBruck:2015tna} that physics of preheating after inflation can be influenced by disformal coupling for this choice of the parameters.
However, we will see in the following section that the disformal coupling terms can drive parametric resonance in the preheating process as long as the term $D (\dot\phi)^2 / C$ is not much smaller than unity although $\gamma \sim 1$.

\section{Preheating in a model with disformal coupling}
\label{sec3}
What we are interested in the present work is to investigate the preheating process after inflation. We will first assume that the spacetime and the inflaton field $\phi$ give a classical background and another scalar field $\chi$ is treated as a quantum field on that background. We also neglect the back reaction of the field $\chi$ on the background field $\phi$. We specialize the Klein-Gordon equations for the fields $\phi$ and $\chi$ to a cosmological background and find respectively
\begin{eqnarray} \label{KGphi}
\left(1+\frac{D}{C}\gamma^{2}\rho_{\chi}\right)\ddot{\phi} +3H\dot{\phi} \left(1-\frac{D}{C}\gamma^{2}\rho_{\chi}\right) +{\cal U}'(\phi) =  \frac{1}{2}\Big[\left(\gamma^{2} - 2\right)\rho_{\chi} +3\gamma^{2}P_{\chi}\Big] \frac{C'}{C} - \frac{1}{2}\left(\gamma^{2} - 1\right)\rho_{\chi}\frac{D'}{D}\,, 
\end{eqnarray}
and
\begin{eqnarray} \label{KGchi}
\ddot{\chi} +3H\dot{\chi} -\frac{1}{\gamma^{2}a^{2}}\nabla^{2}\chi+ \gamma^{2}\frac{D}{C}\dot{\phi}\ddot{\phi}\dot{\chi}+\frac{C}{\gamma^{2}}{\cal V}'(\chi) =  \frac{1}{2}\Big[\left(\gamma^{2} - 3\right)\frac{C'}{C} -\left(\gamma^{2} - 1\right)\frac{D'}{D}\Big]\dot{\phi}\dot{\chi} \,,
\end{eqnarray}
where 
\begin{eqnarray} \label{gamma}
\gamma^{2} = \frac{1}{1-\frac{D}{C}\dot{\phi}^{2}}\,.
\end{eqnarray}
Here, the energy density $\rho_\chi$ and pressure $P_{\chi}$ of the field $\chi$ are respectively given by
\begin{equation}\label{rpchi}
\rho_\chi = \gamma C\left( \frac{1}{2}\gamma^2\dot{\chi}^2+C {\cal V}\right) \,,
\qquad
P_{\chi} = \frac{C}{\gamma} \left( \frac{1}{2}\gamma^2\dot{\chi}^2 - C {\cal V}\right)\,.
 \end{equation}
From the Einstein equation given in the previous section,
we obtain the friedmann equation:
\begin{equation}\label{h2}
3 H^2 = \rho_\chi + \frac 12 (\dot\phi)^2 + {\cal U}(\phi)\,,
\end{equation}
where $H \equiv \dot a / a$ is the Hubble parameter.

\subsection{Inflationary stage}
\label{evo-inf}

In order to examine whether the disformal coupling can influence preheating process, we study the evolution of the inflaton field $\phi$ and $\gamma$ during inflationary stage by firstly assuming that the matter field $\chi$ has no contribution to the dynamics of inflation.
In the case where the contribution from the field $\chi$ can be neglected
and ${\cal U}(\phi)= m^{2}_{\phi}\phi^{2} / 2$, Eq.(\ref{KGphi}) becomes
\begin{eqnarray} \label{KGphia}
\ddot{\phi} +3H\dot{\phi}+m^{2}_{\phi}\phi =  0\,.
\end{eqnarray}
Under the slow-roll evolution, it follows from Eq.(\ref{h2}) that $H \simeq m_\phi \phi / \sqrt{6}$ and therefore the above equation becomes
\begin{equation}\label{slowroll}
\dot\phi \simeq - \sqrt{\frac 23 } m_\phi\,.
\end{equation}
Since $m_\phi^2 \phi^2 \gg \dot\phi^2$ is required for slow-roll evolution, the above equation implies that $\phi > 1$ during inflation. From Eq.(\ref{gamma}),
we see that $\gamma^2$ becomes infinite in the limit when $D \dot\phi^2 / C = 1$.
Thus we have either $\gamma^2 \in [1, \infty)$ or $\gamma^2 \in (-\infty, 0)$ throughout the evolution of the universe.
Nevertheless, $\gamma^2 \in (-\infty, 0)$ is not possible when $\dot\phi$ oscillates around zero during preheating,
and hence we consider only the case $\gamma \geq 1$.
Substituting $\dot\phi$ from Eq.(\ref{slowroll}) into Eq.(\ref{gamma}),
we find that $\gamma$ is larger when $2D m_\phi^2 / (3 C)$ gets closer to unity.
According to the observational bound $m_{\phi} < 10^{-6}$ \cite{ArmendarizPicon:2007iv},
$\gamma$ will significantly larger than unity if $D/C \gg 10^{12}$.
Since we suppose that the dynamics of the universe during inflation is solely governed by the field $\phi$,
the coefficients $C$ and $D$ are not constrained by  observational bounds on inflation.
The upper bound on $\gamma$ is set by $D\gamma^2\rho_\chi / C \ll 1$ to ensure that the contribution from $\chi$ on dynamics of inflation can be neglected. In the following, we quantify the inflationary trajectory in two separate cases.
\subsubsection{{\rm Case I:} $\alpha=\beta$}
Since $\dot\phi$ is nearly constant during inflation,
the simplest situation in which $\gamma$ is large until the end of inflation is that a ratio $D/C$ is constant.
It follows from Eq.(\ref{DC}) that the ratio $D/C$ is constant when $\alpha = \beta$,
which includes the case of constant $C$ and $D$ if $\alpha = \beta = 0$.
Since the universe stops acceleration, i.e., inflation ends,
when $\dot\phi^2 = {\cal U}$,
we find that $\dot\phi$ drops by factor $\sqrt{3} \,\phi_{e} / 2$ at the end of inflation.
Here, $\phi_e \sim 1$ is the value of the inflaton field at the end of inflation.
Hence, we expect that at the end of inflation, the term $D \dot\phi^2 / C$ is in the same order of magnitude with that during inflation.
We will see that this term can contribute to the parametric resonance process.
\subsubsection{{\rm Case II:} $\alpha > \beta\,,\beta<0$}
In addition to the case of a constant $D/C$, a large $\gamma$ at the end of inflation can also be obtained if $D/C$ increases when $\phi$ decreases.
Since $\phi$ decreases near the end of inflation,
the behavior of $D/C$ implies that although $\gamma \sim 1$ during inflation, $\gamma$ can become larger when the value of $\phi$ significantly drops around the end of inflation. This situation corresponds to the case where $\alpha > \beta$, $\beta<0$.

\begin{figure}[t]
\begin{center}		
\includegraphics[width=0.75\linewidth]{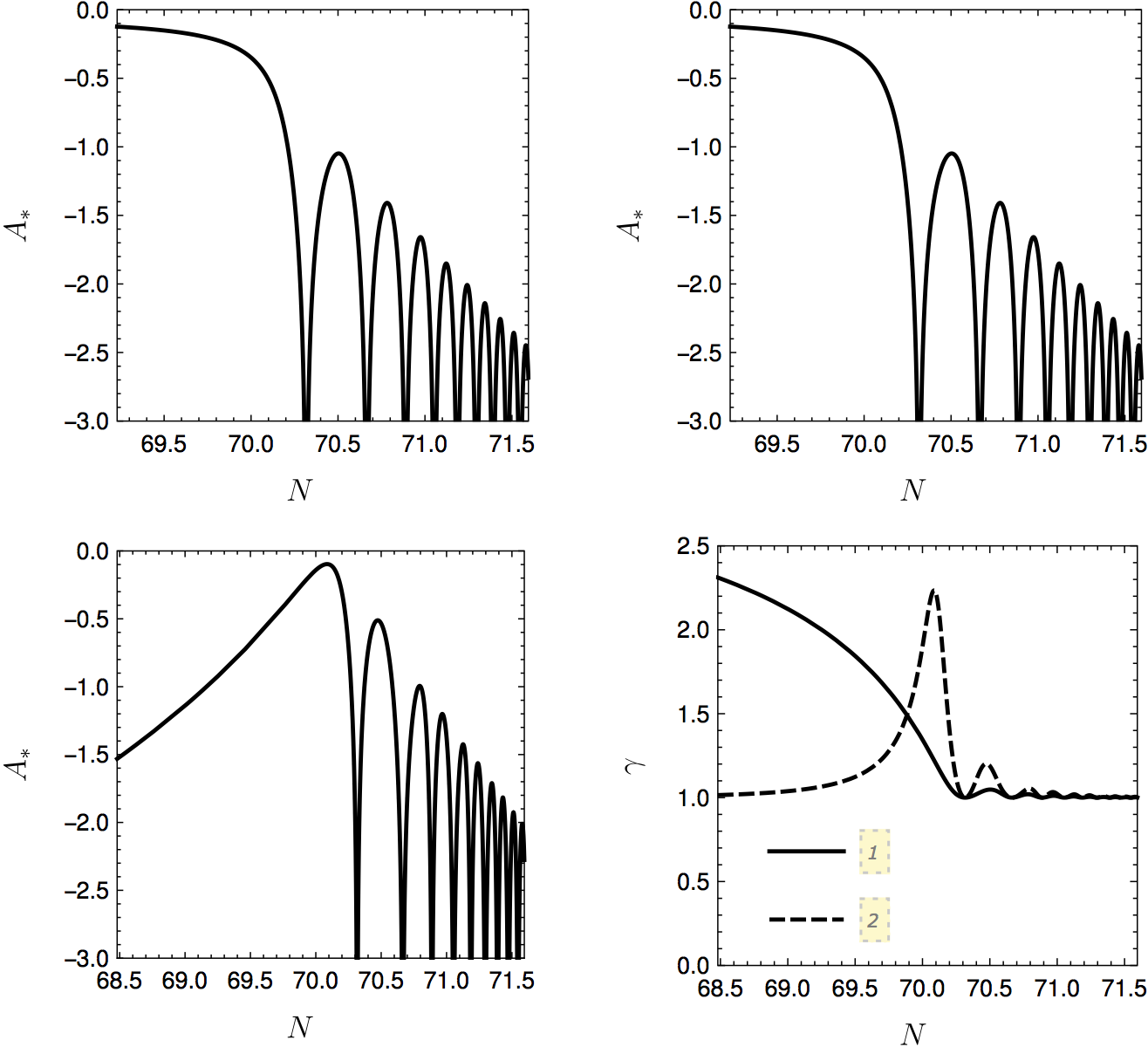}
\caption{\label{fig:evo}
The upper panals show the evolution of $A_* \equiv \log(D \dot\phi^2 /C)$ after inflation for the cases  where $\alpha = \beta = 0$ (left panal) and $\alpha = \beta =1$ (right panal).
The lower left panal shows the evolution of $A_*$ for the case where $(\alpha,\beta) = (0,-2)$.
The lower right panal shows the evolution of $\gamma$.
In this panal, the lines ``1'' and ``2'' represent the cases where $\alpha = \beta =0$ and $(\alpha,\beta) = (0,-2)$, respectively.
In the plots for the case $\alpha=\beta$, we set $C_0 = 1$ and $D_0 = 0.9 (3/2) 10^{14}$,
while we set $C_0 =1$ and $D_0 = 0.9 (3/2) e^{0.06 \phi_i}
10^{14}$ for the case $\alpha \neq \beta$.
}
\end{center}
\end{figure}

After the end of  inflation, $\dot\phi$ oscillates with decreasing amplitude.
To estimate the amplitude of $\gamma$ and $A_{*} \equiv \log(D \dot\phi^2 /C)$ during preheating period for both Case I and Case II,
we examine Eq.(\ref{KGphia}) numerically by setting $m_\phi = 10^{-7}$
and setting  $\phi = \phi_i = \sqrt{276}$ as well as $\dot\phi = - \sqrt{2/3} \,10^{-7}$ at the initial time of inflation. The initial conditions for $\chi$ are set such that $\chi$ has negligible contribution to dynamics of the universe  throughout the inflationary epoch.
In this situation, $\phi$ is required to slowly evolve during inflation, such that $\ddot\phi \ll |H \dot\phi|$ and $H \gg |\dot\phi|$.
Using these conditions and $|D \dot\phi^2 / C| \lesssim 1$,
the terms on the RHS of Eq.(\ref{KGchi}) and the fourth term on the LHS of this equation can be neglected compared with the others.
Hence, during inflation, $\chi$ always slowly evolves if $|C {\cal V}' | \ll |3 H \dot\chi|$,
and evolves as an underdamp oscillation if $|C {\cal V}' | > |3 H \dot\chi|$.
This implies that if $\chi$ initially has no contribution to the dynamics of the universe, its contributions can be neglected  throughout the inflationary epoch without  any special fine-tuning.
When the preheating process starts, the number density of $\chi$ as well as $\rho_\chi$ will be enhanced mainly due to parametric resonances. Bearing in mind that back reaction of $\chi$ to the background field is ignored in our analysis. Based on our consideration, hence, the contributions from $\chi$ to the dynamics of the inflaton field $\phi$ can be neglected during inflation and the first stage of preheating is safe from fine-tuning of the parameters and also initial conditions for $\chi$ are not necessary. The plots of $\gamma$ and $A_{*}$ are shown in Fig.(\ref{fig:evo})
For the selected values of $\alpha$ and $\beta$ in the plots, the chosen value of $D_0$ can lead to largest $\gamma$ at the beginning of preheating.
Increasing $D_0$ from this value cannot significantly enhance  $\gamma$,
while decreasing $D_0$ from this value will suppress magnitude of $\gamma$  during preheating.

\subsection{Preheating stage}
\label{preheat}
Supposing that the process of parametric resonance is at the stage in which the backreaction of the created particle can be neglected. However, as already mentioned in Ref.\cite{GarciaBellido:2008ab}, the backreaction of the quantum field $\chi$ to the dynamics of the inflaton field will be relevant if its occupation numbers have grown sufficiently and then can inhibit the resonance particle production. We will also leave this interesting topic for our future investigation.    

Another word of saying, we are interested in an epoch at which the inflaton is dominant, so that the evolution equation for $\phi$ during preheating is also given by Eq.(\ref{KGphia}). It is trivial to figure out the solution of Eq.(\ref{KGphia}) during preheating. Simply, we use for a power-law evolution of a scale factor $a \propto t^{p}$ and equation of motion becomes:
\begin{eqnarray} \label{KGphiat}
t^{2}\ddot{\phi} + 3tp\dot{\phi}+t^{2}m^{2}_{\phi}\phi =  0\,.
\end{eqnarray}
The general solution of the effective equation of $\phi$ can be basically expressed in terms of the Bessel functions as 
\begin{eqnarray} \label{sophi}
\phi(t) = \frac{1}{(m_{\phi}t)^{u}}\left[AJ_{+u}(m_{\phi}t) + BJ_{-u}(m_{\phi}t)\right]\,,
\end{eqnarray}
where $A$ and $B$ are constants depending on the initial conditions at the end of inflation, and $J_{\pm u}(m_{\phi}t)$ are Bessel functions of order $\pm u$, with $u=(3p-1)/2$. It is well known that the second term of Eq.(\ref{sophi}) diverges in the limit $m_{\phi}t\rightarrow 0$. Therefore, keeping the first term is good enough for the estimate purpose. The physical solution to Eq.(\ref{KGphiat}) is then reduced to:
\begin{eqnarray} \label{sophi1}
\phi(t) = A\,(m_{\phi}t)^{-(3p-1)/2}\,J_{+u}(m_{\phi}t)\,.
\end{eqnarray}
For a large argument expansion of fractional Bessel functions such that $m_{\phi}t\rightarrow 0$, the physical solution can be approximately by a cosinusoidal function, see Ref.\cite{GarciaBellido:2008ab}:
\begin{eqnarray} \label{sophire}
\phi(t) = A\,(m_{\phi}t)^{-(3p-1)/2}\,\cos(m_{\phi}t - 3p\pi/4)\,.
\end{eqnarray}
Here we can choose a constant $A$ by considering the oscillatory behavior which starts just at the end of inflation. As already mentioned above, the field value at the end of inflation is approximately given by $\phi_{e} \sim {\cal O}(1)$. Following Ref.\cite{GarciaBellido:2008ab}, the energy and pressure densities associated to the physical solution (\ref{sophi}) are given, after averaging over several oscillations, by
\begin{eqnarray} \label{omega}
&&\rho_{\phi} \approx \left<\frac{1}{2}{\dot \phi}^{2} + \frac{1}{2}m^{2}_{\phi}\phi^{2}\right> 
\approx \frac{1}{2}m^{2}_{\phi}Z^{2}\left[\left<\sin^{2}(m_{\phi}t - 3p\pi/4)\right> + \left<\cos^{2}(m_{\phi}t - 3p\pi/4)\right>\right] 
= \frac{1}{2}m^{2}_{\phi}Z^{2}\nonumber\\
&&P_{\phi} \approx \left<\frac{1}{2}{\dot \phi}^{2} + \frac{1}{2}m^{2}_{\phi}\phi^{2}\right> 
\approx \frac{1}{2}m^{2}_{\phi}Z^{2}\left[\left<\sin^{2}(m_{\phi}t - 3p\pi/4)\right> - \left<\cos^{2}(m_{\phi}t - 3p\pi/4)\right>\right] 
=0\,,
\end{eqnarray}
with $Z(t) \propto (m_{\phi}t)^{-3p/2}$. Since the averaged pressure is fairly negligible $p_{\phi} \approx 0$, Eq(\ref{omega}) implies that $a(t) \propto t^{p}$ with $p=2/3$. Using the above approximations, the physical solution is finally expressed as
\begin{eqnarray} \label{sophif}
\phi(t) \approx \frac{\phi_{e}}{m_{\phi}t}\,\sin(m_{\phi}t)\,.
\end{eqnarray}
\begin{figure}[t]
\begin{center}		
\includegraphics[width=0.7\linewidth]{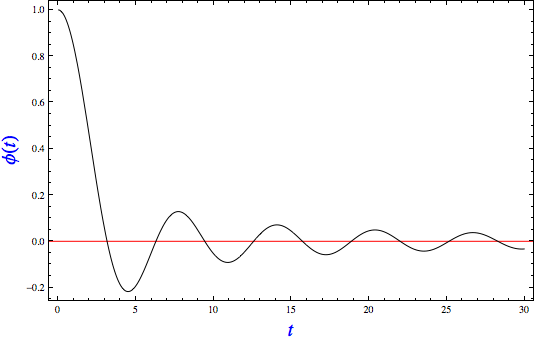}
\caption{We plot the approximate solution of the field $\phi(t)$ as given in Eq.(\ref{sophif}). The value of the scalar field here is measured in units of $M_{\rm P}$ and time is measured in units of $m^{-1}_{\phi}$. \label{plotphi}}
\end{center}
\end{figure}
Typically, it would be of great interesting for evaluating the Hubble constant during the first oscillation. As shown in Fig.(\ref{plotphi}), during the first period of oscillation, the amplitudes of the field $\phi(t)$ drops to around $1/10$ of the reduced Planck mass, $M_{\rm P}$. We may also expect during this early phase of oscillation that the field's kinetic energy is roughly equal to its potential energy, and hence we can estimate the energy density of the field to be $\rho_{\phi} \sim m^{2}_{\phi}\phi^{2}\sim \frac{1}{100}m^{2}_{\phi}M^{2}_{\rm P}$. This approximation allows us to further estimate the Hubble parameter and we find that the Hubble rate would then be $H=\sqrt{\frac{1}{3M^{2}_{\rm P}}\rho}\sim m_{\phi}/\sqrt{300} \sim 0.06\,m_{\phi}$. Notice that the estimate for $H/m_{\phi} \sim 0.06$ is consistent with the results found in Ref.\cite{DeCross:2015uza,Channuie:2016xmq} for completely different scenarios.

We next consider the dynamics of the quantum field $\chi$. Since both C and D are in general functions of a scalar field, we first consider the fourth term of Eq(\ref{KGchi}) and write for an coefficient of $\dot{\chi}$
\begin{eqnarray} \label{chigen}
\gamma^{2}\frac{D}{C}\dot{\phi}\ddot{\phi} &=& \frac{\gamma^{2}}{2}\left(\frac{d}{dt}\Big[\frac{D{\dot{\phi}}^{2}}{C}\Big] + \frac{C'\dot{\phi}}{C}\left(\frac{D{\dot{\phi}}^{2}}{C}\right) - \frac{D'\dot{\phi}}{D}\left(\frac{D{\dot{\phi}}^{2}}{C}\right)\right)\nonumber\\&=& \frac{\gamma^{2}}{2}\left(\frac{d}{dt}\Big[\frac{D{\dot{\phi}}^{2}}{C}\Big] + \frac{C'\dot{\phi}}{C}\left(\frac{\gamma^{2}-1}{\gamma^{2}}\right) - \frac{D'\dot{\phi}}{D}\left(\frac{\gamma^{2} -1}{\gamma^{2}}\right)\right)\nonumber\\&=&\frac{\gamma^{2}}{2}\frac{d}{dt}\Big[\frac{D{\dot{\phi}}^{2}}{C}\Big] + \frac{C'\dot{\phi}}{2C}\left(\gamma^{2}-1\right) -  \frac{D'\dot{\phi}}{2D}\left(\gamma^{2} -1\right)\,.
\end{eqnarray}
After substituting back into Eq.(\ref{KGchi}), we find
\begin{eqnarray} \label{KGchigen}
\ddot{\chi} +3H\dot{\chi} -\frac{1}{\gamma^{2}a^{2}}\nabla^{2}\chi
+ \frac{\gamma^{2}}{2}\frac{d}{dt}\Big[\frac{D{\dot{\phi}}^{2}}{C}\Big]\dot{\chi}+\frac{C}{\gamma^{2}}{\cal V}'(\chi) =  -\frac{C'}{C}\dot{\phi}\dot{\chi} \,.
\end{eqnarray}
The above expression can be further simplified by considering 
\begin{eqnarray} \label{KGchigeng}
3H\dot{\chi} + \frac{\gamma^{2}}{2}\frac{d}{dt}\Big[\frac{D{\dot{\phi}}^{2}}{C}\Big]\dot{\chi}+\frac{C'}{C}\dot{\phi}\dot{\chi} = \frac{d}{dt}\left(\log\Big[\frac{Ca^{3}}{\sqrt{1-\frac{D}{C}\dot{\phi}^{2}}}\Big]\right)\dot{\chi}\,.
\end{eqnarray}
By defining a new parameter:
\begin{eqnarray} \label{newA}
{\cal A}\equiv\Big[\frac{C a^{3}}{\sqrt{1-\frac{D}{C}\dot{\phi}^{2}}}\Big] \,,
\end{eqnarray}
and substituting the above relations Eq.(\ref{KGchigeng}-\ref{newA}) into Eq.(\ref{KGchigen}), we obtain
\begin{eqnarray} \label{KGchigen1}
\ddot{\chi} +{\tilde {\cal H}}\dot{\chi} -\frac{1}{\gamma^{2}a^{2}}\nabla^{2}\chi+\frac{C}{\gamma^{2}}{\cal V}'(\chi) =  0 \,,
\end{eqnarray}
where ${\tilde H}$ is defined as
\begin{eqnarray} \label{newH}
{\tilde {\cal H}}\equiv \frac{\dot{{\cal A}}}{{\cal A}}\,.
\end{eqnarray}
Expanding the scalar fields $\chi$ in terms of the Heisenberg representation as
\begin{eqnarray}
\chi(t,{\bf x}) \sim \int \left(a_{k}\chi_{k}(t)e^{-i{\bf k}\cdot{\bf x}} + a^{\dagger}_{k}\chi^{*}_{k}(t)e^{i{\bf k}\cdot{\bf x}}\right)d^{3}{\bf k} \,, \label{kspace}
\end{eqnarray}
where $a_{k}$ and $a^{\dagger}_{k}$ are annihilation and creation operators, we find that $\chi_{k}$ obeys the following equation of motion:
\begin{eqnarray}
\ddot{\chi}_{k}+{\tilde {\cal H}}\dot{\chi}_{k} +\frac{k^{2}}{\gamma^{2}a^{2}}\chi_{k}+\frac{C}{\gamma^{2}}{\cal V}'(\chi_{k}) = 0\,. \label{kmode}
\end{eqnarray}
Fourier transforming this equation and rescaling the field by $Y_{k} = {\cal A}^{1/2}\chi_{k}$ yields
\begin{eqnarray}
\ddot{Y}_{k} + \left(\frac{k^{2}}{\gamma^{2}a^{2}} 
- \frac 14 \left(2{\dot{\tilde{\cal H}}} + {\tilde{\cal H}}^{2} \right)\right) Y_{k} 
+\frac{C}{\gamma^{2}}{\cal V}' = 0\,.
\label{Ymodegen}
\end{eqnarray}
Setting the potential ${\cal V}(\chi)$ as ${\cal V}(\chi)=m^{2}_{\chi}\chi^{2} / 2$ ,
Eq.(\ref{Ymodegen}) becomes
\begin{eqnarray}
0 &\approx&
\ddot{Y}_{k} + \left(\frac{k^{2}}{a^{2}} +Cm^{2}_{\chi} - \left(C m^{2}_{\chi} + \frac{k^{2}}{a^{2}}\right)\frac{D}{C} {\dot \phi}^{2}
- \frac 14 \left(2 \dot{{\tilde {\cal H}}} + {\tilde {\cal H}}^{2}\right)\right) Y_{k}
\nonumber\\
&\approx&
\ddot{Y}_{k} + \left\{\frac{k^{2}}{a^{2}} +Cm^{2}_{\chi} - \left(C m^{2}_{\chi} + \frac{k^{2}}{a^{2}}\right)\frac{D}{C}{\dot \phi}^{2}
- \frac 14 \left[\underbrace{
\left(3 H + \alpha \dot\phi 
+ \frac{D/C}{1 - D\dot\phi^2/C}F_1
\right)^2
}_{{\tilde {\cal H}}^{2}}
\right.\right.
\nonumber\\
&&\left.\left.
\underbrace{
+ 2\left(3 \dot H + \alpha \ddot\phi
+ 2 \left(\frac{D/C}{1 - D \dot\phi^2 /C}\right)^2F_1^2
+ \frac{D/C}{1 - D \dot\phi^2 / C} \lambda\dot\phi F_1 
+ \frac{D/C}{1 - D \dot\phi^2 / C} \frac{dF_1}{dt} 
\right)
}_{2 \dot{{\tilde {\cal H}}}}
\right]\right\} Y_{k}\,,
\label{DCcont0}
\end{eqnarray}
where $\lambda \equiv \beta - \alpha$ and $F_1 \equiv \lambda \dot\phi^3 / 2 + \dot\phi\ddot\phi$.
Using $a \propto t^{2/3}$ and $H \ll m_{\phi}$ which implies $m_\phi t \gg 1$ during the preheating,
the above equation becomes
\begin{eqnarray}
\ddot{Y}_{k} &+& \left[\frac{k^{2}}{a^{2}} + C m^{2}_{\chi} - \left(C m^{2}_{\chi} + \frac{k^{2}}{a^{2}}\right)\frac{D}{C}{\dot \phi}^{2}
- \frac 12 \alpha m_\phi^2 \phi
- \frac 14 \alpha^2 \dot\phi^2
- \frac 12\frac{D/C}{(1 - D \dot\phi^2/C)} \frac{d}{dt} \left(\dot\phi\ddot\phi\right)\right] Y_{k}
\approx 0\,,
\label{DCcont}
\end{eqnarray}
where we have neglected small contributions coming from terms like ${\cal O}((m_\phi t)^{-3})$ and ${\cal O}((m_\phi t)^{-4})$. It is noticed that, however, $D_0 \dot\phi^2$ is always less than unity  although $D_0 \gg 1$ due to slow-roll evolution during inflation. This quantity can increase in time but cannot be larger than unity after inflation because $\gamma$ becomes infinite when this quantity equals to one. Note that the form of this equation is very similar to that of Ref.\cite{ArmendarizPicon:2007iv} in the context of the derivative coupling between the inflaton and scalar fields.

Surprisingly, preheating in our model may be qualitatively different from the cases considered in various literatures. This is because the field has an approximatedly effective squared mass for $C_{0}=1$ given by
\begin{eqnarray}
m^{2}_{\rm eff.} \approx e^{\alpha\phi} m_\chi^2 - \left(e^{\alpha\phi}m^{2}_{\chi} + \frac{k^{2}}{a^{2}}\right)D_{0} e^{\lambda\phi}{\dot \phi}^{2}
- \frac 12 \alpha m_\phi^2 \phi
- \frac 14 \alpha^2 \dot\phi^2
- \frac 12\frac{D_0 e^{\lambda\phi}}{(1 - D_0 e^{\lambda\phi}\dot\phi^2)} \frac{d}{dt} \left(\dot\phi\ddot\phi\right)\,.
\label{meff}
\end{eqnarray}
By substituting the approximated solution given in Eq.(\ref{sophif}) into the above expression, we get
\begin{eqnarray}
m^{2}_{\rm eff.} &\approx& e^{\alpha\phi}m^{2}_{\chi} - \left(e^{\alpha\phi}m^{2}_{\chi} + \frac{k^{2}}{a^{2}}\right)D_{0}e^{\lambda\phi}m^{2}_{\phi}\left(\frac{\phi_{e}}{m_{\phi}t}\cos(m_{\phi}t)\right)^2
- \frac 12 \alpha m_\phi^2 \frac{\phi_{e}}{m_{\phi}t}\,\sin(m_{\phi}t)
\nonumber\\
&&
- \frac 14 \alpha^2m_\phi^2\left(\frac{\phi_{e}}{m_{\phi}t}\,\cos(m_{\phi}t)\right)^2
+ D_{0} e^{\lambda\phi}m_\phi^4 \left(\frac{\phi_{e}}{m_{\phi}t}\right)^2 
\frac{\cos(2 m_{\phi}t)}
{(1 - D_{0}e^{\lambda\phi} m^{2}_{\phi}\phi_{e}^2 \cos^{2}(m_{\phi}t) / (m_{\phi}t)^{2})}\,,
\label{meffr}
\end{eqnarray}
where we have again neglected small contributions coming from terms like ${\cal O}((m_\phi t)^{-3})$ and ${\cal O}((m_\phi t)^{-4})$. Likewise, it is noticed that the $\lambda$-dependent terms only contribute to the higher orders and can also be neglected in our analysis. It is also worth noting that for $\alpha\neq 0$ the leading-order terms stem from the conformal coupling, while the higher ones come from the disformal coupling. In the situation where $\alpha \neq 0$, the dominant terms in the above expression are
\begin{eqnarray}
m^{2}_{\rm eff.} &\approx& m^{2}_{\chi}  + \alpha\Big[m^{2}_{\chi} - \frac{1}{2}m^{2}_{\phi}\Big]\,\frac{\phi_{e}}{m_{\phi}t}\sin(m_{\phi}t)\,.
\label{meffr1}
\end{eqnarray}
However, for the case with $\alpha = 0$, we instead have
\begin{eqnarray}
m^{2}_{\rm eff.} &\approx& m^{2}_{\chi} - \left(m^{2}_{\chi} + \frac{k^{2}}{a^{2}}\right)D_{0}e^{\lambda\phi}m^{2}_{\phi}\left(\frac{\phi_{e}}{m_{\phi}t}\cos(m_{\phi}t)\right)^2
\nonumber\\
&&
+ D_{0} e^{\lambda\phi}m_\phi^4 \left(\frac{\phi_{e}}{m_{\phi}t}\right)^2 
\frac{\cos(2 m_{\phi}t)}
{(1 - D_{0}e^{\lambda\phi}  m^{2}_{\phi}\phi_{e}^2 \cos^{2}(m_{\phi}t) / (m_{\phi}t)^{2})}\,,
\label{meffr2}
\end{eqnarray}
It follows from the above equation that if $m_\chi \gg m_\phi$, the second line of Eq.(\ref{meffr2}) can be neglected. For simplicity, we will assume that $e^{\lambda\phi} \sim 1$ in the following consideration.
In terms of $m^{2}_{\rm eff.}$, Eq.(\ref{DCcont}) becomes
\begin{eqnarray}
\ddot{Y}_{k} + \omega^{2}_{k}(t) Y_{k} = 0\,,
\label{Mat}
\end{eqnarray}
where a time-dependent frequency of modes $Y_{k}$ is given by
\begin{eqnarray}
\omega^{2}_{k}(t) = m^{2}_{\rm eff.} + \frac{k^{2}}{a^{2}}\,.
\label{fre}
\end{eqnarray}
Certainly, Eq.(\ref{Mat}) describes an oscillatory behavior with a periodically changing frequency $\omega_{k}(t)$. In case of $a=1$, the physical momentum ${\bf p}$ coincides with ${\bf k}$ for Minkowski space such that $k=\sqrt{\bf k}$. The periodicity of Eq.(\ref{Mat}) may in principle lead to the parametric resonance for modes
with certain values of $k$. In order to quantify the parametric resonance behavior in our model, we consider two-separate cases as follows:

\subsubsection{{\rm Parametric resonance for} $\alpha\neq 0\,\,\&\,\,\alpha\geq \beta$}

In this case, $m_{\rm eff.}$ is given by Eq.(\ref{meffr1}). The periodicity of Eq.(\ref{Mat}) for this case may lead to the parametric resonance for modes with certain values of $k$. In order to examine this behavior, we will introduce a new variable, $z$, defined by $m_{\phi}t=2z - \pi/2$. In the Minkowski space for which we neglect the expansion of the universe taking $a=1$ and use a simple trigonometric identity, the equation of motion for the perturbations $Y_{k}$ given in Eq.(\ref{Mat}) can be rewritten in a form of the Mathieu equation, albeit with $q\rightarrow q(t)$:
\begin{eqnarray}
\frac{d^{2}Y_{k}}{dz^{2}} + \left(A_{k} -2q\cos(2z)\right)Y_{k} = 0\,,
\label{Mathe}
\end{eqnarray}
where
\begin{eqnarray}
A_{k} = \left(\frac{2k}{m_{\phi}}\right)^{2} + \left(\frac{2m_{\chi}}{m_{\phi}}\right)^{2}\quad{\rm and}\quad 
q = \frac{2\alpha}{m^{2}_{\phi}}\left(m^{2}_{\chi} - \frac{1}{2}m^{2}_{\phi}\right)\frac{\phi_{e}}{m_{\phi}t}\,.
\label{Mathe1}
\end{eqnarray}
Regarding the Mathieu equation, the solutions are known to exhibit parametric resonance, a.k.a resonance for certain values of the dimensionless parameters of $A_{k}$ and $q$. In the $A_{k}$-$q$ plane, these resonance solutions form band-like patterns called instability bands. Any mode lies along these unstable solutions exhibit exponential growth: $\chi_{k}\propto \exp(\mu\,m_{\phi} t)$ where the characteristic exponent $\mu$ depends on $A_{k}$ and $q$.

To guarantee enough efficiency for the production of particles, the Mathieu equation's parameters should satisfy the broad-resonance conditions, that is $A_{k}\simeq n^{2}$ and $q\gg 1$ where $n$ is an integer. In order for our parameters in Eq.(\ref{Mathe1}) satisfying the broad-resonance conditions, we discover that $m_{\chi}\gg m_{\phi}/\sqrt{2}$ allowing the mode functions lie within at least one of the instability bands and grow exponentially. In this case, we mimic that the effective particle number density in this process increases exponentially. 

Nevertheless, in general, the parameter $q$ can also depend on time and then decreases with time. Therefore it must take a large enough initial value. In the present analysis, we find that there are two cases for this model to initially provide large values for $q$: (I) $m_{\chi}\gg m_{\phi}/\sqrt{2}$ or (II) $\alpha\gg 1$ (if $m_{\chi}\sim m_{\phi}$). Hence, for the disformal coupling scenario, the parametric resonance can proceed efficiently to reheat the universe. This result is different from that obtained in the another derivatively-coupled model \cite{ArmendarizPicon:2007iv}.

\subsubsection{{\rm Parametric resonance for} $\alpha=0\,\,\&\,\,\alpha\geq \beta$}
We now turn to the case when $\alpha =0,\,\,\alpha\geq \beta$ and assume that $e^{\lambda\phi} \sim 1$. Again in order to examine the periodicity behavior of Eq.(\ref{Mat}), we will introduce a new variable, $z$, relating to $m_{\phi}$ via $m_\phi t = 2 z$. Hence the equation of motion for the perturbations $Y_{k}$ given in Eq.(\ref{Mat}) can be recast in a form of the Mathieu equation, albeit with $q\rightarrow q(t)$::
\begin{eqnarray}
\frac{d^{2}Y_{k}}{dz^{2}} + \left(A_{k} -2q\cos(2z)\right)Y_{k} = 0\,,
\label{Mathe2}
\end{eqnarray}
where
\begin{eqnarray}
A_{k} = \left(\frac{2k}{m_{\phi}}\right)^{2} + \left(\frac{2m_{\chi}}{m_{\phi}}\right)^{2}-2q \quad{\rm and}\quad 
q = \frac{1}{m^{2}_{\phi}}\left(m^{2}_{\chi} + \frac{k^{2}}{a^{2}}\right)\frac{D_{0}\phi^{2}_{e}}{t^{2}}\,.
\label{Mathe13}
\end{eqnarray}

To guarantee enough efficiency for the production of particles in this case, we find for the broad-resonance conditions ($q\gg 1$ and $A_{k}\simeq n^{2}$) that the values of $m_{\chi}$ should be much greater that those of $m_{\phi}$. Moreover, in this case, there is no need for $D_{0}$ to be large since the broad resonance can occur when $m_\chi \gg m_\phi$. Moreover, we have another broad-resonance condition such that the field $\chi$ can be light, i.e. $m_{\chi} \lesssim m_{\phi}$. Regarding this condition, we require a large value of $D_{0}$ and the long-wavelenght modes may remain inside the instability band at $A_{k} <0$ allowing the broad resonances in this case can be typically achieved.

\section{Conclusion}

Let us summarize our investigation by first comparing our results with previously studied model with derivative couplings. The study of preheating in derivatively-coupled inflationary models was examined by the authors of Ref.\cite{ArmendarizPicon:2007iv}. In this model, including the expansion of the universe but neglecting back reaction from $\chi$, the mode functions $\chi_{k}$ and the homogenous inflaton field $\phi$ can be combined into the Mathieu equation, albeit with $q\rightarrow q(t)$, and the parameters $A_{k}$, and $q$ take the form
\begin{eqnarray}
A_{k} = \frac{k^{2}}{m^{2}_{\phi}a^{2}} + \frac{m^{2}_{\chi}}{m^{2}_{\phi}} -2q \quad{\rm and}\quad q = \frac{1}{2}\left(\frac{\phi_{e}}{F\,t}\right)^{2}\,,\label{ndc}
\end{eqnarray}
with $F$ being a vacuum expectation value of the inflaton field. As mentioned in Ref.\cite{ArmendarizPicon:2007iv}, an initial value of $q$ is more efficient for longer wavelength modes and lighter fields. Notice from Eq.(\ref{ndc}) that a value of $q$ is at most of order one since $\phi_{e} \approx M_{\rm P} \approx F$. If this is the case, any resonance proceeds close to the end of instability bands is ineffective if the field is very heavy, $m_{\chi}\gg m_{\phi} (A_{k}\gg 1)$. 

Contrary to previously studied model with derivative couplings presented in Ref.\cite{ArmendarizPicon:2007iv}, we have shown above that the parametric resonance in our model is rather effective with certain conditions.  We demonstated that for $\alpha\neq 0\,\,\&\,\,\alpha\geq \beta$ our parameters in Eq.(\ref{Mathe1}) satisfy the broad-resonance conditions if (I) $m_{\chi}\gg m_{\phi}/\sqrt{2}$ or (II) $\alpha\gg 1$ (if $m_{\chi}\sim m_{\phi}$). For $\alpha=0\,\,\&\,\,\alpha\geq \beta$, to guarantee enough efficiency for the production of particles, the values of $m_{\chi}$ should be much greater that those of $m_{\phi}$. Moreover, if the field $\chi$ be light, i.e. $m_{\chi} \lesssim m_{\phi}$, the broad resonances exhibit the instability band at $A_{k} <0$. We have also found that for the disformal coupling model considered here,
the dominant contribution to the preheating process come from conformal coupling if the conformal coefficient is time-dependent, i.e., $\alpha \neq 0$.

However, we have neglected the back reactions of the field $\chi$ to the background field $\phi (t)$ and reduce the behavior of the background field to that of a simple, minimally coupled field. Otherwise, this situation is much more complicated.  Moreover, regarding the reference \cite{ArmendarizPicon:2007iv,Tsujikawa:1999iv}, we anticipated our present analysis not to constitute an impasse for our model.

\acknowledgments
The work of K. Karwan is financially supported by the Thailand Research Fund (TRF) with Grant No. RSA5780053.

\end{document}